\documentclass[aps,a4paper,superscriptaddress,showpacs,preprintnumbers,amsmath,amssymb]{revtex4}

\usepackage{ulem}

\usepackage{psfrag} \usepackage{graphicx} \usepackage{dcolumn}
\usepackage{color} \usepackage{latexsym,amsfonts} \usepackage{bm}
\usepackage{amssymb}
\begin{document}
\setcounter{page}{1}
\def\theequation{\arabic{section}.\arabic{equation}}
\setcounter{section}{0}

\title{Storage rings as detectors for relic gravitational-wave
  background ?}

\author{A. N. Ivanov}\email{ivanov@kph.tuwien.ac.at}
\affiliation{Atominstitut, Technische Universit\"at Wien, Stadionallee
  2, A-1020 Wien, Austria}
\author{\fbox{A. P. Kobushkin}}\email{kobushkin@bitp.kiev.ua}
\affiliation{Bogoliubov Institute for Theoretical Physics, 03143, Kiev
  and Physical and Technical National University KPI, Prospect Pobedy
  37, 03056, Kiev, Ukraine}
\author{M. Wellenzohn}\email{max.wellenzohn@gmail.com}
\affiliation{Atominstitut, Technische Universit\"at Wien, Stadionallee
  2, A-1020 Wien, Austria} \affiliation{FH Campus Wien, University of
  Applied Sciences, Favoritenstra\ss e 226, 1100 Wien, Austria}

\date{\today}

\begin{abstract}
We argue that storage rings can be used for the detection of
low-frequency gravitational-wave background.  We explain the
systematic shrinkage of the machine circumference of the storage ring
of the SPring-8, observed by Takao and Shimada (Proceedings of EPAC
2000, Vienna, 2000, p.1572), by the influence of the relic
gravitational-wave background.  We show that the forces, related to
the {\it stiffness} of the physical structures, governing the path of
the beam, can be neglected for the analysis of the shrinkage of the
machine circumference caused by the relic gravitational-wave
background. We show that the systematic shrinkage of the machine
circumference can be explained by a relic gravitational-wave
background even if it is treated as a stochastic system incoming on
the plane of the machine circumference from all quarters of the
Universe. We show that the rate of the systematic shrinkage of the
machine circumference does not depend on the radius of the storage
ring and it should be universal for storage rings with any radius.
\end{abstract}

\maketitle

\section{Introduction}
\setcounter{equation}{0}
\label{sec:introduction}

The existence of gravitational waves has been predicted by Einstein
\cite{Einstein1916, Einstein1917} within Einstein's general theory of
relativity \cite{Einstein1916GR} (see also \cite{Misner1973,
  Peacock1999}). Starting with the pioneering work by Weber
\cite{Weber1960} one of the most challenging problems of experimental
physics is the detection of gravitational radiation.  In the seventies
of the last century the existence of gravitational waves has been
confirmed indirectly in a set of accurate measurements of secular
orbital period changes in the Hulse-Taylor binary pulsar
\cite{Taylor1982, Taylor1989}. An attempt for the observation of the
cosmic low-frequency gravitational-wave background has been undertaken
by Stinebring {\it et al.}  \cite{Stinebring1990}. Recently
\cite{Abbott2016}, LIGO and Virgo scientific collaborations have
succeeded in detecting the gravitational waves in the frequency range
100 - 300\, {\rm Hz}. Such a success of LIGO and Virgo scientific
collaborations might provide a background for the existence of
gravitational waves in very different frequency ranges.

An interesting influence of gravity on the variation of the machine
circumference of the storage rings has been found in
\cite{Arnaudon1995, Date1999, Takao2000}. As has been shown by
Arnaudon {\it et al.}  \cite{Arnaudon1995} the machine circumference
of the Large Electron Positron Collider (LEP) at CERN is sensitive to
the tidal forces, induced by a motion of the Moon around the Earth. An
analogous influence of a motion of the Moon around the Earth on the
variations of the machine circumference of the storage ring of the
SPring-8 (the Super Photon Ring of 8\,{\rm GeV} by RIKEN in Japan) has
been observed by Dat\'e and Kugamai \cite{Date1999}. In turn, Takao
and Shimada \cite{Takao2000} have found that the machine circumference
of the storage ring of the SPring-8 undergoes itself not only an
influence of the tidal forces, caused by a motion of the Moon around
the Earth, but also the seasonal forces, induced by a rotation of the
Earth around the Sun. However, Takao and Shimada \cite{Takao2000}, in
addition to the seasonal variations of the machine circumference of
the storage ring of the SPring-8, have detected a steady damping of
the machine circumference of an unknown nature. Below we discuss only
the measurements for the SPring-8 storage ring by Takao and Shimada
\cite{Takao2000}, though our results should be, in principle,
applicable also to any storage rings.

The paper is addressed to the analysis of the steady damping or the
systematic shrinkage of the machine circumference of the SPring-8,
observed by Takao and Shimada \cite{Takao2000}. Since on the Earth the
main fundamental effect of the gravitaitonal waves is the {\it
  gravitational strain}, i.e. a fractional distortion in the lengths
of the objects induced by the gravitational waves \cite{Peacock1999},
we assume that the observed steady damping of the machine
circumference of the storage ring of the SPring-8 is caused by an
influence of the gravitational waves with sufficiently low frequencies.
Thus, a total change of the machine circumference of the SPring-8 we
propose to define as follows
\begin{eqnarray}\label{label1.1}
\frac{\Delta C}{\Delta t} = \Bigg(\frac{\Delta C}{\Delta t}\Bigg)_{\rm
m} + \Bigg(\frac{\Delta C}{\Delta t}\Bigg)_{\rm s} +
\Bigg(\frac{\Delta C}{\Delta t}\Bigg)_{\rm gw},
\end{eqnarray}
where first two terms are caused by the tidal (m) and seasonal (s)
forces, whereas the third term describes a rate of a steady damping of
the machine circumference, caused by the gravitational waves.

The paper is organized as follows. In section \ref{sec:strain} we
estimate the rate of the shrinkage of the machine circumference due to
the gravitational strain. In section \ref{sec:equation} we solve the
equations of motion of the storage ring in the field of the
cylindrical relic gravitational wave. We show that the solution of the
equations of motion gives the same result obtained within the
hypothesis of the {\it gravitational strain}. We show that the rate of
the shrinkage of the machine circumference does not depend on the
radius of the storage ring and should be universal for storage rings
with any radii.  In section \ref{sec:stiffness} we
discuss the influence of the {\it stiffness} of the physical
structures of the storage ring, governing the path of the beam. We
argue that the forces, induced by the {\it stiffness} of the physical
structures of the storage ring, governing the path of the beam, can be
neglected for the analysis of the shrinkage of the machine
circumference caused by the relic gravitational-wave background. In
Section 6 we investigate the shrinkage of the machine circumference
induced by a stochastic spherical relic gravitational-wave background
incoming on the plane of the machine circumference from all quarters
of the Universe. We show that the stochastic relic gravitational-wave
background, incoming on the plane of the machine circumference from
all quarters of the Universe, does not destroy the shrinkage of the
machine circumference. The former is due to the fact that the effect
of the shrinkage of the machine circumference is of the second order
in gravitational wave interactions. We show that the independence of
the rate of the shrinkage of the machine circumference on the radius
of the storage ring retains in the case of the interaction of the
storage ring with the stochastic relic gravitational-wave
background. In the Conclusion we discuss the obtained results.

\section{Gravitational strain and  shrinkage of machine circumference}
\setcounter{equation}{0}
\label{sec:strain}

According to the {\it gravitational strain} as a fundamental effect
of the action of gravitational waves on the lengths of different
objects on the Earth, we assume that the variation $\delta C_{\rm
  gw}(t)$ of the machine circumference with the length $C_0$ is
induced by low-frequency gravitational waves , which we treat as
perturbations of the space-time metric with a signature
$(-1,+1,+1,+1)$ \cite{Misner1973, Peacock1999, Kostelecky2004}.  For
the calculation of $\delta C_{\rm gw}(t)$ we define a perturbation of
the metric $h_{ab}(t,z)\,(a,b = x,y,z)$ as a monochromatic plane wave
traveling along the $z$-axis with a frequency $\omega$ and a momentum
$k = \omega$ \cite{Misner1973, Peacock1999}
\begin{eqnarray}\label{eq:2.1}
h_{ab}(t,z) = \left(\begin{array}{llcl} h_{xx}(t,z) & h_{xy}(t,z) &
0\\ h_{yx}(t,z) & h_{yy}(t,z) & 0\\ ~~0 & ~~0 & 0
\end{array}\right) = \left(\begin{array}{llcl}
\Delta_+ & ~\Delta_\times & 0\\
 \Delta_\times & - \Delta_+ & 0\\
~0 & ~~0 & 0
\end{array}\right)\,\cos(\omega t - k z + \delta),
\end{eqnarray}
where $\Delta_+$ and $\Delta_\times$ are constant amplitudes of the
diagonal and non-diagonal components of the monochromatic plane wave,
$h_{tt}(t,z) = h_{ta}(t,z) = h_{at}(t,z) = 0$ \cite{Misner1973,
  Peacock1999} and $\delta$ is an arbitrary phase. We define the
monochromatic plane wave in the so-called {\it transverse traceless}
gauge $h_{aa}(t,z) = h_{xx}(t,z) + h_{yy}(t,z) = 0$ (see pp.946-948
of Ref.\cite{Misner1973}).

Placing the storage ring in the $xy$-plane at $z = 0$ the variation
$\delta C_{\rm gw}(t)$ can be defined by the contour integral
\begin{eqnarray}\label{eq:2.2}
\delta C_{\rm gw}(t) &=& \oint_{C_0}\sqrt{dx^2 + dy^2 +
  h_{xx}(t,0)dx^2 + h_{yy}(t,0)dy^2 + 2h_{xy}(t,0)dxdy} -
C_0=\nonumber\\ &=& \frac{C_0}{2\pi}\int^{2\pi}_0\Big(\sqrt{1 -
  h_{xx}(t,0)\cos2\varphi - h_{xy}(t,0)\sin2\varphi} - 1\Big)\,d\varphi =
- \frac{1}{16}\,C_0\,(h^2_{xx}(t,0) + h^2_{xy}(t,0)) =\nonumber\\ &=&
- \frac{1}{16}\,C_0\,h^2_0\,\cos^2(\omega t + \delta),
\end{eqnarray}
where we have used polar coordinates $x = (C_0/2\pi)\cos\varphi$ and
$y = (C_0/2\pi)\sin\varphi$ and the relation $h_{yy}(t,0) = -
h_{xx}(t,0)$ (\ref{eq:2.1}) and the amplitude $h_0$ is equal to $h_0 =
\sqrt{\Delta^2_+ + \Delta^2_\times}$. We would like to accentuate that
the amplitude $h_0$ of the monochromatic plane wave is not the real
amplitude of the relic gravitational-wave background. The relation of
the amplitude $h_0$ to the amplitude $h^{\rm gw}_0$ of the relic
gravitational-wave background can be found, for example, as
follows. Indeed, a real relic gravitational-wave background should be
treated as a perturbation of the Friedmann-Lema\^itre-Robertson-Walker
(FLRW) metric \cite{Misner1973, Peacock1999}
\begin{eqnarray}\label{eq:2.3}
ds^2 = - dt^2 + a^2(t)\Big(\frac{dr^2}{1 - kr^2} + r^2\big(d\theta^2 +
\sin^2\vartheta\,d\varphi^2\big) + a^2(t)\,r^2\,\cos^2\vartheta\big(-
h^{\rm gw}_{xx}\,\cos 2\varphi - h^{\rm gw}_{xy}\,\sin
2\varphi\big)d\varphi^2 + \ldots\Big),
\end{eqnarray}
where $a(t)$ is the expansion parameter and $k = 0, \pm 1$ stands for
flat, closed and open Universe, respectively \cite{Misner1973,
  Peacock1999}. Defining the locations of the material points in the
FLRW space-time in the coordinate system related to the coordinate
system of the storage ring of the SPring-8, for the change of the
machine circumference of the storage ring $\delta C_{\rm gw}(t)$ we get the following expression
\begin{eqnarray}\label{eq:2.4}
&&\delta C_{\rm gw}(t) = - \frac{1}{16}\,R_{\rm U}\,(h^{(\rm
    gw)}_0)^2\,\cos^2(\omega t + \delta),
\end{eqnarray}
where $ R_{\rm U} \sim a(t_{\rm U}) = 4.4\times 10^{26}\,{\rm m}$ is
the radius of the observable Universe at the present time $t_{\rm U} =
13.80(2)\,{\rm Gyr}$ after the Big Bang \cite{PDG2020}.

We would like to notice that it is rather clear that the contribution
of the gravitational waves to the variation of the machine
circumference, $\delta C_{\rm gw}(t) \sim O((h^2_0)$, is of the second
order.  Indeed, the mass quadrupole moment of the storage ring,
located in the $xy$-plane at $z = 0$, has only two equal components
$D_{xx} = D_{yy} = D$. Due to this, the interaction of this mass
quadrupole moment with gravitational waves is proportional to
$D\,(h_{xx} + h_{yy})$, which is zero by definition for gravitational
waves in the {\it transverse traceless gauge} $h_{xx} = - h_{yy}$
\cite{Misner1973, Peacock1999}.

The change of the storage ring of the machine circumference $\Delta
C_{\rm gw}$ induced by the gravitational waves (\ref{eq:2.1}) for the
time interval $\Delta t = t_2 - t_1$ is equal to
\begin{eqnarray}\label{eq:2.5}
\Delta C_{\rm gw} = \delta C_{\rm gw}(t_2) - \delta C_{\rm gw}(t_1) =
\frac{1}{16}\,C_0\,h^2_0\,\sin(\omega \Delta t)\,\sin(2 \omega t +
2\delta).
\end{eqnarray}
where $t = (t_2 + t_1)/2$.  For the rate of the change of the machine
circumference at $\Delta t \to 0$ we get
\begin{eqnarray}\label{eq:2.6}
\frac{\Delta C_{\rm gw}(t)}{\Delta t} =
\frac{1}{16}\,C_0\,h^2_0\,\omega\,\sin(2\omega t + 2\delta).
\end{eqnarray}
For the comparison with the experimental rate we have to average the
theoretical rate (\ref{eq:2.6}) over the data-taking period
$\tau$. This gives
\begin{eqnarray}\label{eq:2.7}
\hspace{-0.3in}\Big\langle\frac{\Delta C_{\rm gw}(t)}{\Delta
t}\Big\rangle_{\tau} =
\frac{1}{16}\,C_0\,h^2_0\,\omega\,\frac{1}{\tau}
\int^{+\tau/2}_{-\tau/2}dt\,\sin(2\omega t + 2\delta) =
\frac{1}{16}\,C_0\,h^2_0\,\sin 2\delta\;\frac{\sin \omega \tau}{\tau}.
\end{eqnarray}
In the low-frequency limit $\omega \tau \ll 1$, corresponding to the
case of the relic gravitational-wave background, the relation
(\ref{eq:2.5}) can be transcribed into the form 
\begin{eqnarray}\label{eq:2.8}
\frac{1}{C_0}\Big\langle\frac{\Delta C_{\rm gw}(t)}{\Delta
t}\Big\rangle_{\tau} = \frac{1}{16}\,h^2_0\,\omega\,\sin 2\delta .
\end{eqnarray}
Since the experimental rate of the change of the machine circumference
is equal to
\begin{eqnarray}\label{eq:2.9}
 \frac{1}{C_0}\Big(\frac{\Delta C}{\Delta t}\Big)_{\exp} \simeq -
 1.4\times 10^{-7}\,{\rm yr^{-1}},
\end{eqnarray}
a comparison of theoretical and experimental rates leads to the
relation
\begin{eqnarray}\label{eq:2.10}
\omega\, h^2_0\,\sin 2\delta   = \frac{16}{C_0}\,\Big(\frac{\Delta
C}{\Delta t}\Big)_{\exp} \simeq -\,2.2 \times 10^{-4}\,{\rm yr}^{-1}.
\end{eqnarray}
The experimentally observed shrinkage of the machine circumference of
the storage ring \cite{Takao2000} imposes a constraint on the phase of
the gravitational wave, i.e. $-\sin 2\delta > 0$. Setting for further
estimates $|\sin 2\delta| \sim 1$, we get the frequency of the
gravitational waves $\omega \sim 2.2\times 10^{-6}\,h^{-2}_0\,{\rm
  Gyr^{-1}}$.  This corresponds to an oscillation period $T_{\rm gw} =
2\pi/\omega \sim 3\times 10^{-3}\,h^2_0{\rm Gyr}$ of the shrinkage of
the machine circumference, which is smaller compared with the age of
the Universe $T_{\rm gw} \ll T_{\rm U} \simeq 13.8\,{\rm Gyr}$
\cite{PDG2020}. Since the oscillation period exceeds greatly any
reasonable interval of experimental measurements, the rate of the
shrinkage of the machine circumference, induced by the relic
gravitational-wave background, should be constant in time during any
data-taking period. This agrees with Eq.(\ref{eq:2.8}).

The rate of the shrinkage of the machine circumference, represented in
terms of the relic gravitational-wave perturbations of the FLRW metric
(\ref{eq:2.5}), reads
\begin{eqnarray}\label{eq:2.11}
\Big\langle\frac{\Delta C_{\rm gw}(t)}{\Delta t}\Big\rangle_{\tau} =
\frac{\pi}{8}\,R_{\rm U}\,(h^{\rm gw}_0)^2\,\omega\,\sin 2\delta.
\end{eqnarray}
From the comparison Eqs.(\ref{eq:2.8}) and (\ref{eq:2.11}) we define $h^{\rm gw}_0$ in terms of $h_0$
\begin{eqnarray}\label{eq:2.12}
h^{\rm gw}_0 = \sqrt{\frac{C_0}{2\pi R_{\rm U}}}\,h_0 = 7.2\times
10^{-13}\,h_0,
\end{eqnarray}
where $C_0 = 1436\,{\rm m}$ is the length of the machine circumference
of the storage ring of the SPring-8 \cite{Date1999, Takao2000},

Another important quantity characterizing the relic gravitational-wave
background is the energy-density parameter $\Omega_{\rm gw}$ defined
by
\begin{eqnarray}\label{eq:2.13}
\Omega_{\rm gw} = \frac{\omega^2 (h^{\rm gw}_0)^2}{24 H^2_0} =
\frac{C_0}{2\pi R_{\rm U}}\,\frac{\omega^2h^2_0}{24 H^2_0} = 2.2
\times 10^{-17}\,h^{-2}_0,
\end{eqnarray}
calculated for $\omega h^2_0 = 2200\,{\rm Gyr}^{-1}$ and $H_0 =
6.893\times 10^{-2}\,{\rm Gyr^{-1}}$ \cite{PDG2020}. The estimate of
the amplitude $h_0$ we carry out in section \ref{sec:stiffness}.

\section{Equations of motion for shrinkage of machine circumference}
\setcounter{equation}{0}
\label{sec:equation}

In this section we show that the analysis of the influence of the
relic gravitational-wave background through the solution of equations
of motion for the storage ring in the cylindrical relic
gravitational-wave field gives the same result that we have obtained
in section \ref{sec:strain}. According to \cite{Misner1973} (see
pp.1004-1011 of Ref.\cite{Misner1973}), the non-relativistic motion of
a massive particle in the $xy$-plane at $z = 0$ induced by the
cylindrical gravitational-wave background can be described the
equations of motion
\begin{eqnarray}\label{eq:3.1}
\frac{d^2x}{dt^2}&=& - R^x{}_{0x0}\,x - R^x{}_{0y0}\,y,\nonumber\\
\frac{d^2y}{dt^2}&=& - R^y{}_{0x0}\,x - R^y{}_{0y0}\,y,
\end{eqnarray}
where $R^{\alpha}_{\beta\gamma\delta}$ is the Riemann tensor defined
by \cite{Kostelecky2004}
\begin{eqnarray}\label{eq:3.2}
R^{\alpha}{}_{\mu \beta \nu} = \frac{\partial
  \Gamma^{\alpha}{}_{\beta\mu}}{\partial x^{\nu}} - \frac{\partial
  \Gamma^{\alpha}{}_{\nu\mu}}{\partial x^{\beta}} +
\Gamma^{\alpha}{}_{\nu\varphi} \Gamma^{\varphi}{}_{\beta\mu} -
\Gamma^{\alpha}{}_{\beta\varphi} \Gamma^{\varphi}{}_{\nu\mu}
\end{eqnarray}
The Christoffel symbols or differently the ``covariant connection
coefficients'' $\Gamma^{\alpha}_{\lambda\mu}$ are determined in terms
of the metric tensor \cite{Misner1973}
\begin{eqnarray}\label{eq:3.3}
\Gamma^{\alpha}{}_{\lambda\mu} =
\frac{1}{2}\,g^{\alpha\nu}\,\Big(\frac{\partial g_{\mu\nu}}{\partial
  x^{\lambda}} + \frac{\partial g_{\lambda\nu}}{\partial x^{\mu}} -
\frac{\partial g_{\mu\lambda}}{\partial x^{\nu}}\Big).
\end{eqnarray}
For the calculation of the Christoffel symbols we use the following
metric tensor \cite{Misner1973} 
\begin{eqnarray}\label{eq:3.4}
g_{\mu\nu} = \eta_{\mu\nu} + h_{\mu\nu}(t - z),
\end{eqnarray}
where $\eta_{\mu\nu} = {\rm diag}(- 1, 1, 1, 1)$ and $h_{\mu\nu}(t -
z)$ is a symmetric tensor defined in the {\it transverse traceless
gauge} with non-zero components $h_{xx}(t - z) = - h_{yy}(t - z)$ and
$h_{xy}(t - z) = h_{yx}(t - z)$ \cite{Misner1973}.  In terms of
$h_{\mu\nu}(t - z)$ the Christoffel symbols are given by
\begin{eqnarray}\label{eq:3.5}
\Gamma^{\alpha}{}_{\lambda\mu} =
\frac{1}{2}\,\eta^{\alpha\nu}\,\Big(\frac{\partial
h_{\mu\nu}}{\partial x^{\lambda}} + \frac{\partial
h_{\lambda\nu}}{\partial x^{\mu}} - \frac{\partial
h_{\mu\lambda}}{\partial x^{\nu}}\Big) -
\frac{1}{2}\,h^{\alpha\nu}\,\Big(\frac{\partial h_{\mu\nu}}{\partial
x^{\lambda}} + \frac{\partial h_{\lambda\nu}}{\partial x^{\mu}} -
\frac{\partial h_{\mu\lambda}}{\partial x^{\nu}}\Big).
\end{eqnarray}
The components of the Riemann tensor contributing to the equations of
motion (\ref{eq:3.1}) are given by
\begin{eqnarray}\label{eq:3.6}
R^x{}_{0x0} &=& \frac{d\Gamma^x{}_{0x}}{dt} + \Gamma^x{}_{x0}\Gamma^x{}_{0x}
+ \Gamma^x{}_{y0}\Gamma^y{}_{0x}\quad,\quad R^x{}_{0y0} = 
\frac{d\Gamma^x{}_{0y}}{dt} + \Gamma^x{}_{x0}\Gamma^x{}_{0y} +
\Gamma^x{}_{y0}\Gamma^y{}_{0y},\nonumber\\ R^y{}_{0x0} &=& 
\frac{d\Gamma^y{}_{0x}}{dt} + \Gamma^y{}_{x0}\Gamma^x{}_{0x} +
\Gamma^y{}_{y0}\Gamma^y{}_{0x}\quad,\quad R^y{}_{0y0} = 
\frac{d\Gamma^y{}_{0y}}{dt} + \Gamma^y{}_{x0}\Gamma^x{}_{0y} +
\Gamma^y{}_{y0}\Gamma^y{}_{0y}.
\end{eqnarray}
The Christoffel symbols read
\begin{eqnarray}\label{eq:3.7}
\hspace{-0.3in}\Gamma^x{}_{0x}&=& + \frac{1}{2}\,\frac{dh_{xx}}{dt} -
\frac{1}{4}\, \frac{d}{dt}(h^2_{xx} + h^2_{xy})\quad\quad\;,\;
\Gamma^y{}_{0y} = - \frac{1}{2}\,\frac{dh_{xx}}{dt} - \frac{1}{4}\,
\frac{d}{dt}(h^2_{xx} + h^2_{xy}),\nonumber\\
\hspace{-0.3in}\Gamma^x{}_{0y}&=& + \frac{1}{2}\,\frac{dh_{xy}}{dt} -
\frac{1}{2}\,h_{xx} h_{xy}\,\frac{d}{dt}{\ell
  n}\Big(\frac{h_{xy}}{h_{xx}}\Big)\;,\; \Gamma^y{}_{0x} = +
\frac{1}{2}\,\frac{dh_{xy}}{dt} + \frac{1}{2}\,h_{xx}
h_{xy}\,\frac{d}{dt}{\ell n}\Big(\frac{h_{xy}}{h_{xx}}\Big).\nonumber\\
\hspace{-0.3in}&&
\end{eqnarray}
For the monochromatic gravitational waves the ratio $h_{xy}/h_{xx} =
\Delta_{\times}/\Delta_+$ is constant and the Christoffel symbols
$\Gamma^x{}_{0y}$ and $\Gamma^y{}_{0x}$ are linear in $h_{ab}$. For
the calculation of the components of the Riemann tensor, defining the
equations of motion (\ref{eq:3.1}), we keep also the terms of order
$O(h^2_{ab})$ inclusively and obtain
\begin{eqnarray}\label{eq:3.8}
R^x{}_{0x0} &=&  + \frac{1}{2}\,\frac{d^2h_{xx}}{dt^2} -
\frac{1}{4}\,\frac{d^2}{dt^2}(h^2_{xx} + h^2_{xy}) +
\frac{1}{4}\,\Big[\Big(\frac{dh_{xx}}{dt}\Big)^2 +
\Big(\frac{dh_{xy}}{dt}\Big)^2\Big],\nonumber\\ R^x{}_{0y0} &=&
R^y{}_{0x0} =  \frac{1}{2}\,\frac{d^2h_{xy}}{dt^2} ,\nonumber\\
R^y{}_{0y0} &=&- \frac{1}{2}\,\frac{d^2h_{xx}}{dt^2} -
\frac{1}{4}\,\frac{d^2}{dt^2}(h^2_{xx} + h^2_{xy}) +
\frac{1}{4}\,\Big[\Big(\frac{dh_{xx}}{dt}\Big)^2 +
\Big(\frac{dh_{xy}}{dt}\Big)^2\Big].
\end{eqnarray}
Substituting (\ref{eq:3.8}) in the equations of motion
(\ref{eq:3.1}) we get
\begin{eqnarray}\label{eq:3.9}
\ddot{x}&=& - \frac{1}{2}\,\ddot{h}_{xx}(t)\,x -
\frac{1}{2}\,\ddot{h}_{xy}(t)\,y + \frac{1}{4}\,(\ddot{h^2}(t) -
\dot{h}^2(t))\,x, \nonumber\\ \ddot{y}&=&
- \frac{1}{2}\,\ddot{h}_{xy}(t)\,x + \frac{1}{2}\,\ddot{h}_{xx}(t)\,y +
\frac{1}{4}\,(\ddot{h^2}(t) - \dot{h}^2(t))\,y,
\end{eqnarray}
where over-dots stand for the derivative with respect to time. We have
denoted $h^2 = h^2_{xx} + h^2_{xy}$ and $\dot{h}^2 = \dot{h}^2_{xx} +
\dot{h}^2_{xy}$. The equations of motion (\ref{eq:3.9}) can be treated
as the Lagrange equations derived from the Lagrange function
\begin{eqnarray}\label{eq:3.10}
\hspace{-0.3in}L(t,x,y,\dot{x},\dot{y}) = \frac{1}{2}\,(\dot{x}^2 +
\dot{y}^2) + \frac{1}{8}\,(\ddot{h^2}(t) - \dot{h}^2(t))\,(x^2 + y^2)-
\frac{1}{4}\,\ddot{h}_{xx}(t)\,(x^2 - y^2) -
\frac{1}{2}\,\ddot{h}_{xy}(t)\,xy.
\end{eqnarray}
In the polar coordinates $x = r\cos\Phi$ and $y = r\sin \Phi$ we
get
\begin{eqnarray}\label{eq:3.11}
\hspace{-0.3in}L(t, r, \varphi, \dot{r}, \dot{\Phi}) =
\frac{1}{2}\,\dot{r}^2 + \frac{1}{8}\,(\ddot{h^2}(t) -
\dot{h}^2(t))\,r^2 + \frac{1}{2}\,r^2\dot{\Phi}^2 -
\frac{1}{4}\,r^2\,(\ddot{h}_{xx}(t)\,\cos 2\Phi +
\ddot{h}_{xy}(t)\,\sin 2\Phi).
\end{eqnarray}
Assuming that the radius $r$ is almost constant we can factorize
radial and angular degrees of freedom. 
\begin{eqnarray}\label{eq:3.12}
  \hspace{-0.21in}L(t, r, \varphi, \dot{r}, \dot{\Phi}) =
  \frac{1}{2}\,\dot{r}^2 - \frac{1}{8}\,\dot{h}^2(t)\,r^2 +
  R^2_0\,\Big[\frac{1}{2}\,\dot{\Phi}^2 -
    \frac{1}{4}\,(\ddot{h}_{xx}(t)\,\cos 2\Phi +
    \ddot{h}_{xy}(t)\,\sin 2\Phi)\Big].
\end{eqnarray}
where $R_0$ is the radius of the machine circumference, $R_0 =
C_0/2\pi$. In such an approximation the term
$(1/8)\,\ddot{h^2}\,R^2_0$ is a total derivative, which can be removed
from the Lagrange function. The equations of motion for the radius
$r(t)$ and the azimuthal angle $\Phi(t)$ are equal to
\begin{eqnarray}\label{eq:3.13}
\ddot{r}(t) = - \frac{1}{4}\, \dot{h}^2(t)\,R_0 \quad,\quad
\ddot{\Phi}(t) = \frac{1}{2}\,(\ddot{h}_{xx}(t)\,\sin 2\Phi(t) -
\ddot{h}_{xy}(t)\,\cos 2\Phi(t)).
\end{eqnarray}
Since $\Phi(t) \ll 1$, the solution to Eq.(\ref{eq:3.13}) for the
azimuthal motion reads
\begin{eqnarray}\label{eq:3.14}
\Phi(t) = - \frac{1}{2}\,h_{xy}(t) = - 
\frac{1}{2}\,\Delta_{\times}\,\cos(\omega t + \delta)\quad,\quad
\dot{\Phi}(t) =  - \frac{1}{2}\,\dot{h}_{xy}(t) = +
\frac{1}{2}\,\Delta_{\times}\,\omega\,\sin(\omega t + \delta).
\end{eqnarray}
For the frequencies of the gravitational wave background corresponding
to the low-frequency limit $\omega \to 0$ we get
\begin{eqnarray}\label{eq:3.15}
\dot{\Phi}(t) =  \frac{1}{2}\,\Delta_{\times}\,\omega\,\sin\delta.
\end{eqnarray}
This predicts a very weak rotation of the machine circumference with a
practically constant velocity in dependence of the polarization and
phase of the gravitational wave background. From (\ref{eq:3.13})
we obtain the rate of the radius of the machine circumference
\begin{eqnarray}\label{eq:3.16}
\frac{1}{R_0}\,\frac{dr(t)}{dt} = - \frac{1}{4}\,\int
dt\,\dot{h}^2(t))
\end{eqnarray}
For the relic monochromatic cylindrical gravitational wave $h_{xx} =
\Delta_+\,\cos(\omega t + \delta)$ and $h_{xy} =
\Delta_\times\,\cos(\omega t + \delta)$ the r.h.s. of
(\ref{eq:3.15}) is equal to
\begin{eqnarray}\label{eq:3.17}
\frac{1}{R_0}\,\frac{dr(t)}{dt} = \frac{1}{16}\,h^2_0\,\omega\,
\sin(2\omega t + 2\delta) - \frac{1}{8}\,h^2_0\omega^2 t.
\end{eqnarray}
To leading order in the low-frequency limit $\omega \to 0$ we get
\begin{eqnarray}\label{eq:3.18}
\frac{1}{C_0}\frac{dC(t)}{dt} = \frac{1}{16}\,h^2_0\,\omega\,\sin
2\delta,
\end{eqnarray}
where $C_0 = 2\pi R_0$ and $dC(t) = 2\pi r(t)$, respectively.  This
agrees fully with our result (\ref{eq:2.8}) obtained within the
hypothesis of the {\it gravitational strain}.

\section{Shrinkage of the machine circumference and stiffness of 
physical structures, governing the path of the beam}
\setcounter{equation}{0}
\label{sec:stiffness}

In this section we discuss the influence of the forces, related to the
{\it stiffness} of the physical structures of the storage ring,
governing the path of the beam (mounts of magnets, for instance) and
estimate the amplitude $h_0$. Indeed, one can imagine that the forces,
induced by the {\it stiffness} of the physical structures of the
storage ring, can prevent the machine circumference of the storage
ring from the shrinkage caused by the relic gravitational-wave
background. Below we show that the forces, caused by the {\it
  stiffness} of the physical structures of the storage ring, can be
neglected for the analysis of the shrinkage of the machine
circumference, induced by the relic gravitational-wave background.

Let $\vec{F}_{\rm stiff}$ be the force, caused by the {\it stiffness}
of the physical structures of the storage ring.  The observation
of the fluctuations of the machine circumference, induced by the tidal
and seasonal forces \cite{Date1999,Takao2000}, assumes that the
forces, produced by the {\it stiffness} of the physical structures of
the storage ring, are smaller compared with the tidal and seasonal
forces.

Since the seasonal forces are smaller compared to the tidal forces but
have been measured experimentally by the change of the machine
circumference, it is obvious that the forces, induced by the {\it
  stiffness} of the physical structures of the storage ring, should be
smaller compared with the seasonal forces. This can be written as
follows $|\vec{F}_{\rm stiff}| \ll | \vec{F}_{\rm s}(\vec{r}\,)|$,
where $\vec{F}_{\rm s}(\vec{r}\,)$ is the seasonal force defined by
\cite{Date1999}
\begin{eqnarray}\label{eq:4.1}
\vec{F}_{\rm s}(\vec{r}\,) = - \bigtriangledown\,U_{\rm s}(\vec{r}\,).
\end{eqnarray}
The potential $U_{\rm s}(\vec{r}\,)$, produced by the Sun, is given by
\cite{Date1999}
\begin{eqnarray}\label{eq:4.2}
U_{\rm s}(\vec{r}\,) = G_N M_{\odot}\,\bigg(\frac{1}{|\vec{R}_s -
  \vec{r}\,|} - \frac{1}{R_s} - \frac{\vec{r}\cdot
  \vec{R}_s}{R^3_s}\bigg),
\end{eqnarray}
where $G_N = 6.636 \times 10^4\,{\rm m^3\,kg^{-1}\,yr^{-2}}$
\cite{PDG2020}, $M_{\odot} = 1.989\times 10^{30}\,{\rm kg}$ is the mass
of the Sun, $R_s = 1.496\times 10^{11}\,{\rm m}$ is the distance
between centers of the Sun and the Earth, $|\vec{r}| = R_{\oplus} =
6.378\times 10^6\,{\rm m}$ is the radius of the Earth.

The rate of the change of the machine circumference, caused by the
tidal and seasonal forces is of order of $|\Delta C/\Delta t| =
4\times 10^{-4}\,{\rm m/yr}$ \cite{Date1999,Takao2000}. The the
experimental rate of the steady damping of the machine circumference,
$|\Delta C/\Delta t| = 2\times 10^{-4}\,{\rm m/yr}$ \cite{Takao2000},
is of the same order of magnitude. This implies that the forces,
leading to the shrinkage of the machine circumference, can be of
gravitational nature. Moreover, the forces, induced by the {\it
  stiffness} of the physical structures of the storage ring, governing
the path of the beam, should be smaller compared to the forces
responsible for the steady damping.

In order to get a quantitative confirmation of this assertion we
suggest to compare the energy densities of the seasonal forces and the
relic gravitational-wave background.  Following \cite{Misner1973,
  Peacock1999} we define the energy density of the gravitational-wave
background and the seasonal forces. The energy-density $\rho_{\rm gw}
= \langle T^{\rm gw}_{00}(t)\rangle$ is defined by
\begin{eqnarray}\label{eq:4.3}
\rho_{\rm gw} = \langle T^{\rm gw}_{00}(t)\rangle = \frac{1}{16\pi
  G_N}\langle \big(R^x{}_{0x0} + R^y{}_{0y0}\big)\rangle =
\frac{1}{32\pi G_N}\langle \big(\dot{h}^2 -\ddot{h^2}\big) \rangle =
\frac{\omega^2 (h^{\rm gw}_0)^2}{64\pi G_N},
\end{eqnarray}
where the components of the Riemann tensor are defined in
Eq.(\ref{eq:3.8}) with a replacement $h_{ab}(t) \to h^{\rm
  gw}_{ab}(t)$. The energy density of the seasonal forces $\rho_s =
\langle T^s_{00}(t)\rangle$ is defined by
\begin{eqnarray}\label{eq:4.4}
\rho_s = \langle T^s_{00}(t)\rangle = \frac{1}{16\pi
  G_N}\langle \big(R^x{}_{0x0} + R^y{}_{0y0} + R^z{}_{0z0}\big)\rangle.
\end{eqnarray}
The components of the Riemann tensor in Eq.(\ref{eq:4.4}) we calculate
by taking into account that the potential $U_s(\vec{r}\,)$ is a part
of the metric tensor. Since on the surface of the Earth $\vec{r} =
\vec{R}_{\oplus}$, we have
\begin{eqnarray}\label{eq:4.5}
- g_{00} = 1 + 2 U_s(\vec{R}_{\oplus})\,\cos(\omega_s t +
    \delta_s)\quad,\quad g_{ab} = \Big(1 - 2
    U_s(\vec{R}_{\oplus})\,\cos(\omega_s t + \delta_s)\Big)\,
    \delta_{ab}.
\end{eqnarray}
During experiments by Takao and Shimada \cite{Takao2000} a periodicity
of the seasonal force was caused by a motion of the Earth around the
Sun with a period $T_s = 2\pi/\omega_s = 0.5\,{\rm yr}$. We take into
account such a periodicity by the function $\cos(\omega_s t +
\delta_s)$ and average over time. As a result we obtain
\begin{eqnarray}\label{eq:4.6}
\rho_s = \langle T^s_{00}(t)\rangle  = \frac{~~3 \omega^2_s}{32 \pi
      G_N}\langle U^2_{\rm s}(\vec{R}_{\oplus})\rangle =\frac{~~3
      \omega^2_s G_N M^2_{\odot}}{32 \pi
    }\Big\langle\Big(\frac{3(\vec{R}_{\oplus}\cdot \vec{R}_s)^2 -
      R^2_{\oplus}R^2_s}{2 R^5_s}\Big)^2\Big\rangle = \frac{ 3 \pi}{40}\,
    \frac{G_N M^2_{\odot} R^4_{\oplus}}{T^2_s R^6_s},
\end{eqnarray}
where we have integrated over directions of the vector $\vec{R}_s$,
corresponding to a rotation of the Earth around the Sun. Assuming that
$\rho^{\rm gw} \sim \rho^{\rm s}$, we obtain one more relation between
$h^{\rm gw}_0$ and $h_0$
\begin{eqnarray}\label{eq:4.7}
h^{\rm gw}_0 \sim \sqrt{\frac{6}{5}}\, \frac{G_N
    M_{\odot} R^2_{\oplus} T_{\rm gw}}{R^3_s T_s} \simeq
  10^{-9}\,h^2_0, 
\end{eqnarray}
calculated for $T_{\rm gw} \simeq 3\times 10^{-3}\,h^2_0\, {\rm
  Gyr}$. From the comparison of Eqs.(\ref{eq:4.7}) and (\ref{eq:2.12})
we get $h_0 \sim 7\times 10^{-4}$ and $h^{\rm gw}_0 \sim 5\times
10^{-16}$, respectively. The frequency $\omega$ (or the circular
frequency) of the relic gravitational-wave background is of order
$\omega \sim 10^{-7}\,{\rm Hz}$. It corresponds to the linear
frequency $f = \omega/2\pi \sim 2 \times 10^{-8}\,{\rm Hz}$. For such
an estimate the energy-density parameter $\Omega_{\rm gw}$ of the
relic gravitational-wave background, given in Eq.(\ref{eq:2.13}), is
of order $\Omega_{\rm gw} \sim 5 \times 10^{-11}$.

The obtained results should testify that the {\it stiffness} of the
physical structures of the storage ring, governing the path of the
beam, can be neglected for the analysis of the shrinkage of the
machine circumference, caused by the relic gravitational-wave
background.

\section{Stochastic relic gravitational-wave background}
\setcounter{equation}{0}
\label{sec:sgw}

In this section we analyze the systematic shrinkage of the machine
circumference \cite{Takao2000} coupled to the relic gravitational-wave
background treated as a stochastic system \cite{Buonanno2003,
  Regimbau2011, Caprini2018}. We show that the suggested explanation
of the shrinkage of the machine circumference, measured at the
SPring-8 \cite{Takao2000}, by the relic gravitational-wave background
survives even if the storage ring interacts with the stochastic relic
gravitational waves coming from all quarters of the Universe. This is
related to the fact that the observed shrinkage of the machine
circumference of the storage ring is an effect of the second order of
the interaction of the gravitational waves with the machine
circumference.

Below we consider spherical relic gravitational waves \cite{Fock1959}
converging to the center of the machine circumference\,\footnote{The
  spherical gravitational waves, converging to the center of the
  machine circumference, we define as \cite{Fock1959}:
  $h_{ab}(t,|\vec{r} - \vec{R}_0|) \sim \cos(\omega(t - |\vec{r} -
  \vec{R}_0|/c) + \delta)/|\vec{r} - \vec{R}_0|$, where the vector
  $\vec{R}_0$ is the radius-vector of the machine circumference,
  located in the plane of the machine circumference $|\vec{R}_0| = R_0
  = C_0/2\pi$, and $\vec{r}$ is the radius-vector of the observer. It
  is zero , $\vec{r} = 0$, at the center of the machine
  circumference.}. The relic gravitational waves are polarized in the
$(\varphi_s\vartheta_s)$ plane, defined by unit vectors
$\vec{e}_{\varphi_s}$ and $\vec{e}_{\vartheta_s}$ as it is shown in
Fig.1, perpendicular to the direction of the propagation, which is
anti-parallel to the unit vector $\vec{e}_r$. The polarization tensor,
determined in the {\it transverse traceless gauge}, has the following
non-vanishing components: $\Delta_{\varphi_s\varphi_s} = -
\Delta_{\vartheta_s\vartheta_s} = \Delta_+$ and
$\Delta_{\varphi_s\vartheta_s} = \Delta_{\vartheta_s\varphi_s} =
\Delta_\times$.

The {\it gravitational strain} of the machine circumference, induced
by the stochastic relic gravitational waves incoming from all quarters
of the Universe, can be defined by
\begin{eqnarray}\label{eq:5.1}
\hspace{-0.3in}\delta C_{\rm gw}(t) = -  C_0\, \frac{1}{16\pi}\int
d\Omega_s \int^{2\pi}_0\big\langle\big( h_{xx}(t)\,\cos 2\varphi +
h_{xy}(t)\sin2\varphi\big)^2\big\rangle d\varphi,
\end{eqnarray}
where $h_{xx}(t)$ and $h_{xy}(t)$ are given by
\begin{eqnarray}\label{eq:5.2}
h_{xx}(t) &=& \cos\vartheta_s\,(\Delta_+\,\cos 2\varphi_s +
\Delta_\times\,\sin 2\varphi_s)\,\cos\Big(\omega \Big(t -
\frac{R_0}{c}\Big) + \delta\Big),\nonumber\\ h_{xy}(t)&=&
\cos\vartheta_s\,(- \Delta_+\,\sin 2\varphi_s + \Delta_\times\,\cos
2\varphi_s)\,\cos\Big(\omega \Big(t - \frac{R_0}{c}\Big) +
\delta\Big)
\end{eqnarray}
and $h_{yy}(t) = - h_{xx}(t)$. Below we neglect $R_0/c = 7.6\times
10^{-7}\,{\rm s}$, where $R_0 \simeq 229\,{\rm m}$ is the radius of
the machine circumference, relative to the data-taking period $\tau$,
which is about a few years \cite{Takao2000}.  The quantities $h_{xx}(t)$,
$h_{yy}(t)$ and $h_{xy}(t)$ are the projections of the components of
the polarization tensor of the spherical relic gravitational wave on
the plane of the machine circumference (see Fig.1). They depend on the
angles $\vartheta_s$ and $\varphi_s$, which are the angle of the slope
of the polarization plane of the gravitational wave relative to the
plane of the machine circumference and the azimuthal angle,
respectively (see Fig.1). At $\vartheta_s = \varphi_s = 0$ we get a
gravitational wave equivalent to the cylindrical gravitational wave
defined by (\ref{eq:2.1}). 

Integration over the angles $\vartheta_s$ and $\varphi_s$, where
$d\Omega_s = \sin\vartheta_sd\vartheta_sd\varphi_s$, takes into
account the contributions of the stochastic relic gravitational waves
incoming on the plane of the machine circumference from all quarters
of the Universe\,\footnote{It is assumed that the Earth is transparent
  for the relic gravitational waves.}. Below we assume that the
stochastic relic gravitational-wave background is isotropic
\cite{Shea2004}.
\begin{figure}
\centering \psfrag{p}{$\vec{e}_r$} \psfrag{g}{$\vec{e}_{\vartheta_s}$}
\psfrag{a}{$\varphi_s$}
\psfrag{b}{$\vartheta_s$}\psfrag{d}{$\vec{e}_{\varphi_s}$}
\psfrag{x}{$\vec{e}_x$} \psfrag{y}{$\vec{e}_y$}
\psfrag{z}{$\vec{e}_z$}
\includegraphics[height=0.30\textheight]{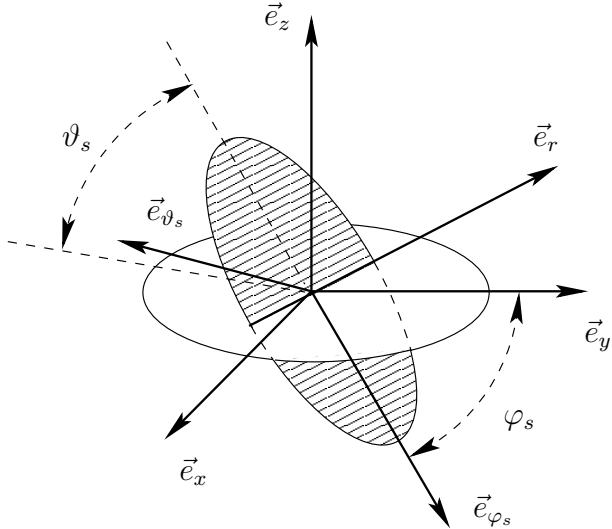}
\caption{The orientation of the polarization plane of the stochastic
  relic gravitational waves relative to the plane of the machine
  circumference.}
\end{figure}
In Eq.(\ref{eq:5.1}) the brackets $\langle \ldots\rangle$ mean
\begin{eqnarray}\label{eq:5.3}
\langle f\rangle = \int^{\infty}_0 d\omega\,S_h(\omega)\,f(\omega),
\end{eqnarray}
where $S_h(\omega)$ is a spectral density, caused by the averaging
over stochastic degrees of freedom of the relic gravitational-wave
background. We suppose that the spectral density $S_h(\omega)$ is
normalized to unity. Plugging (\ref{eq:5.2}) into (\ref{eq:5.1}) we
get
\begin{eqnarray}\label{eq:5.4}
\hspace{-0.3in}\delta C_{\rm gw}(t) &=& -
\frac{C_0}{16\pi}\int^{\infty}_0d\omega\,S_h(\omega)\,\cos^2(\omega t
+ \delta)\int d\Omega_s \,\cos^2\vartheta_s\nonumber\\
\hspace{-0.3in}&\times& \int^{2\pi}_0\,\Big[\Delta_+\,\cos 2(\varphi_s
+ \varphi) + \Delta_\times\,\sin 2(\varphi_s +
\varphi)\Big]^2\,d\varphi,
\end{eqnarray}
Integrating over the angular variables we obtain the {\it
gravitational strain}, induced by the stochastic spherical relic
gravitational-wave background incoming on the plane of the machine
circumference from all quarters of the Universe. It reads
\begin{eqnarray}\label{eq:5.5}
\hspace{-0.3in}\delta C_{\rm gw}(t) &=& -
\frac{4\pi}{3}\,\frac{1}{16}\,C_0 \, h^2_0\int^{\infty}_0
d\omega\,S_h(\omega)\,\cos^2(\omega t + \delta).
\end{eqnarray}
The relative rate of the shrinkage of the machine circumference can be
defined by
\begin{eqnarray}\label{eq:5.6}
\frac{1}{C_0}\,\frac{\Delta C_{\rm gw}}{\Delta t} =
\frac{4\pi}{3}\,\frac{1}{16}\,h^2_0 \int^{\infty}_0
d\omega\,\omega\,S_h(\omega)\,\sin(2\omega t + 2\delta).
\end{eqnarray}
For the comparison of the theoretical rate of the change of the
machine circumference (\ref{eq:5.6}) with the experimental data one
has to average the theoretical rate over the data-taking period
$\tau$. This gives
\begin{eqnarray}\label{eq:5.7}
\frac{1}{C_0}\,\Big\langle \frac{\Delta C_{\rm gw}}{\Delta
t}\Big\rangle_{\tau} &=& \frac{4\pi}{3}\,\frac{1}{16}\,h^2_0
\int^{\infty}_0 d\omega\,\omega\,S_h(\omega)\frac{1}{\tau}\int^{+
\tau/2}_{-\tau/2}dt\,\sin(2\omega t + 2\delta) =\nonumber\\
&=&\frac{4\pi}{3}\,\frac{1}{16}\,h^2_0\,\sin 2\delta \int^{\infty}_0
d\omega\,\omega\,S_h(\omega)\,\frac{\sin \omega \tau}{\omega \tau}.
\end{eqnarray}
Of course, the result of the evaluation of the integral over $\omega$
depends on the shape of the energy density $S_h(\omega)$, which is
model-dependent \cite{Buonanno2003}.

In turn, the rate of the shrinkage of the machine circumference, given in
terms of the stochastic relic gravitational-wave perturbations of the
FLRW metric (\ref{eq:2.5}), is given by
\begin{eqnarray}\label{eq:5.8}
\Big\langle \frac{\Delta C_{\rm gw}}{\Delta t}\Big\rangle_{\tau} =
\frac{4\pi}{3}\,\frac{\pi}{8}\,R_{\rm U}\,(h^{\rm gw}_0)^2\, \sin
2\delta \int^{\infty}_0 d\omega\,\omega\,S_h(\omega)\,\frac{\sin
  \omega \tau}{\omega \tau}
\end{eqnarray}
From the comparison of (\ref{eq:5.8}) with
(\ref{eq:5.8}) one can conclude that the relation between the
amplitudes $h_0$ and $h^{\rm gw}_0$, given by Eq.(\ref{eq:2.12}), is
preserved for the stochastic relic gravitational-wave background
incoming on the plane of the machine circumference from all quarters
of the Universe.

\section{Conclusion}

The results obtained in paper should be understood as a hint that
experimental analysis of fine variations of the machine circumferences
of the storage rings can, in principle, contain an information about
the relic gravitational-wave background on the same footing as the
storage rings are sensitive to the tidal and seasonal forces
\cite{Arnaudon1995}-\cite{Takao2000}.

We argue that if the systematic shrinkage (or a steady damping) of the
machine circumference of the storage ring, observed at the SPring-8
\cite{Takao2000}, is caused by the influence of the relic
gravitational-wave background, the same effect should be, in
principle, observed for the machine circumference of the storage ring
of any accelerator, for example, the LEP and LHC at CERN, the ELSA at
University of Bonn, the DAPHNE at Frascati, the VEPP-4 at Novosibirsk
and others.

We have shown that the rate of the shrinkage of the machine
circumference, represented in terms of the relic gravitational-wave
perturbations of the FLRW metric, does not depend on the length of the
machine circumference and should be universal for any storage ring
with any radius (see Eqs.(\ref{eq:2.11}) and (\ref{eq:5.8})).

This makes very simple the experimental analysis of the validity of
our hypothesis of the influence of the relic gravitational-wave
background on the shrinkage of the machine circumference of the
SPring-8 storage ring. Indeed, it is sufficient to measure the rates
of the shrinkage of the machine circumferences of the storage rings of
the LEP at CERN, the DAPHNE at Frascati, the VEPP-4 at Novosibirsk or
of any other accelerators. If the rates of the shrinkage of the
machine circumferences of the storage rings would have been found
comeasurable with the value $ (\Delta C(t)/\Delta t)_{\exp} =
-\,2\times 10^{-4}\,{\rm m/yr}$, obtained for the SPring-8 storage
ring \cite{Takao2000}, this should testify the detection of the relic
gravitational-wave background. Any negative result should bury the
hypothesis. From the comparison of the energy densities of the
gravitational field, caused by the seasonal forces, and of the relic
gravitational-wave background, we have estimated the parameters
characterizing the relic gravitational-wave background responsible for
the systematic shrinkage of the machine circumference of the storage
ring of the SPring-8. At $- \sin 2\delta \sim 1$ we have got i) $h_0
\sim 7 \times 10^{-4}$, where $h_0$ is the amplitude of the
gravitational-wave perturbation of the metric of the space, related to
the storage ring, ii) $h^{\rm gw}_0 \sim 5 \times 10^{-16}$, where
$h^{\rm gw}_0$ is the amplitude of the relic graviational-wave
background as a perturbation of the FLRW metric, iii) $\omega \sim
10^{-7}\,{\rm Hz}$, $T_{\rm gw} = 2\pi/\omega \sim 2\,{\rm yr}$ and
$\Omega_{\rm gw} \sim 5 \times 10^{-11}$, where $\omega$, $T_{\rm gw}$
and $\Omega_{\rm gw}$ are the circular frequency, period and
energy-density parameter of the relic gravitational-wave
background,inducing a systematic shrinkage of the machine
circumference of the storage ring of the SPring-8.

Finally we have discussed the interaction of the storage ring with a
stochastic relic spherical gravitational-wave background. We have
shown that, since the shrinkage of the machine circumference is a
phenomenon of the second order in gravitational wave interactions, it
cannot be destroyed even if one takes into account the contribution of
the relic gravitational waves incoming on the plane of the machine
circumference from all quarters of the Universe. We have obtained an
additional factor $4\pi/3$ relative to the rate of the shrinkage of
the machine circumference, induced by the cylindrical relic
gravitational-wave background (\ref{eq:2.8}). Of course, the estimates
of the parameters of the relic gravitational-wave background, inducing
a systematic shrinkage of the machine circumference, depend of the
shape of the energy density $S_h(\omega)$, which is model-dependent
\cite{Buonanno2003, Regimbau2011, Caprini2018}.

One can assume that in case of the validity of our explanation of the
systematic shrinkage of the machine circumference, observed by Takao
and Shimada \cite{Takao2000} for the SPring-8, by the relic
gravitational-wave background, the constraints on the averaged
frequencies of the relic gravitational waves can be likely used to set
limits on a low-frequency cosmological spectrum \cite{Buonanno2003,
  Regimbau2011, Caprini2018}. It is also able to place some
constraints on the effective-field-theory extensions of Einstein's
gravitational theory \cite{Sennett2020}.

\section{Acknowledgement}

The work of A. N. Ivanov was supported by the Austrian ``Fonds zur
F\"orderung der Wissenschaftlichen Forschung'' (FWF) under contracts
P31702-N27 and P33279-N.  The work of M. Wellenzohn was supported by
the MA 23 (FH-Call 16) under the project ``Photonik -
Stiftungsprofessur f\"ur Lehre''. All results of this manuscript have
been delivered online at ARIES WP6 Workshop: Storage Rings and
Gravitational Waves "SRGW2021", held from 2.02.2021 to 31.03.2021 and
organized by CERN, Geneva Switzerland.

\end{document}